\documentclass[apj]{emulateapj}

\def\xmm {\emph{XMM-Newton}}

\def\swi {\emph{Swift}}
\def\integral {\emph{INTEGRAL}}
\def\src {1E\,1547.0--5408}

\begin{document}

\title{The dust-scattering X-ray rings of the anomalous X-ray pulsar \src}
\author{A.~Tiengo,\altaffilmark{1} G.~Vianello,\altaffilmark{1} P.~Esposito,\altaffilmark{1,2} S.~Mereghetti,\altaffilmark{1} A.~Giuliani,\altaffilmark{1} E.~Costantini,\altaffilmark{3} G.~L.~Israel,\altaffilmark{4} L.~Stella,\altaffilmark{4} R.~Turolla,\altaffilmark{5,6}
S.~Zane,\altaffilmark{6} N.~Rea,\altaffilmark{7,8} D.~G\"otz,\altaffilmark{9}  F.~Bernardini,\altaffilmark{4} A.~Moretti,\altaffilmark{10} P.~Romano,\altaffilmark{11} M.~Ehle,\altaffilmark{12} and N.~Gehrels\altaffilmark{13}}

\altaffiltext{1}{INAF/Istituto di Astrofisica Spaziale e Fisica Cosmica - Milano, via E.~Bassini 15, 20133 Milano, Italy; tiengo@iasf-milano.inaf.it}
\altaffiltext{2}{INFN - Istituto Nazionale di Fisica Nucleare, Sezione di Pavia,
via A.~Bassi 6, 27100 Pavia, Italy}
\altaffiltext{3}{SRON, Netherlands Institute for Space Research, Sorbonnelaan 2, 3584~CA Utrecht, The Netherlands}
\altaffiltext{4}{INAF/Osservatorio Astronomico di Roma, via Frascati 33, 00040 Monteporzio Catone, Italy}
\altaffiltext{5}{Universit\`a degli Studi di Padova, Dipartimento di Fisica, via F.~Marzolo 8, 35131 Padova, Italy}
\altaffiltext{6}{University College London, Mullard Space Science Laboratory, Holmbury St. Mary, Dorking, Surrey RH5 6NT, UK}
\altaffiltext{7}{Institut de Ci\`encies de l'Espai (CSIC--IEEC), Campus UAB, Facultat de Ci\`encies, Torre C5-parell, 08193 Barcelona, Spain}
\altaffiltext{8}{University of Amsterdam, Astronomical Institute Anton Pannekoek, Kruislaan 403, 1098~SJ Amsterdam, The Netherlands}
\altaffiltext{9}{CEA Saclay, DSM/Irfu/Service d'Astrophysique, Orme des Merisiers, B\^at. 709, 91191 Gif-sur-Yvette, France}
\altaffiltext{10}{INAF/Osservatorio Astronomico di Brera, via E.~Bianchi 46, 23807 Merate, Italy}
\altaffiltext{11}{INAF/Istituto di Astrofisica Spaziale e Fisica Cosmica - Palermo, via U.~La Malfa 153, 90146 Palermo, Italy}
\altaffiltext{12}{\xmm\ Science Operations Centre, ESAC, ESA, P.O. Box 78, 28691 Villanueva de la Ca\~{n}ada, Spain}
\altaffiltext{13}{NASA Goddard Space Flight Center, Greenbelt, Maryland 20771, USA}

\shortauthors{A. Tiengo et al.}
\shorttitle{The distance of the AXP \src\ from its X-ray scattering halo}
\journalinfo{This is an author-created, un-copyedited version of an article accepted for publication in The Astrophysical Journal.}

\submitted{Received 4 November 2009; accepted 18 December 2009}

\begin{abstract}

On 2009 January 22
numerous strong bursts  were detected from the anomalous X-ray pulsar \src.
\swi/XRT and \xmm/EPIC observations carried out in the following two weeks led to the discovery of three X-ray rings centered on this source. The ring radii increased with time following the expansion law expected for a short impulse of X-rays
scattered by three dust clouds.
Assuming different models for the dust composition and grain size distribution, we fit the intensity decay of each ring as a function of time at different energies, obtaining tight constrains on the distance of the X-ray source. Although the distance strongly depends on the adopted dust model, we find that some models are incompatible with our X-ray data, restricting to 4--8 kpc the range of possible distances for \src.
The best-fitting dust model provides a source distance of $3.91\pm0.07$ kpc, which is compatible with the proposed association with the supernova remnant G\,327.24--0.13, and implies distances of 2.2 kpc, 2.6 kpc and 3.4 kpc for the dust clouds, in good agreement with the dust distribution inferred by CO line observations towards \src. However,
dust distances in agreement with CO data are also
obtained for a set of similarly well-fitting models that imply a source distance of $\sim$5 kpc. A distance of $\sim$4--5 kpc is also favored by the fact that these dust models are already known to provide good fits to the dust-scattering halos of bright X-ray binaries.
Assuming $N_{\rm H}=10^{22}$ cm$^{-2}$ in the dust cloud responsible for the brightest ring and a bremsstrahlung spectrum with $kT=100$ keV, we estimate that the burst producing the X-ray ring released an energy of
10$^{44-45}$ erg in the 1--100 keV band,
suggesting that this burst was the brightest flare without any long-lasting pulsating tail ever detected from a magnetar.

\end{abstract}

\keywords{dust, extinction --- stars: neutron --- X-rays: individual (\src) --- X-rays: stars}

\section{Introduction}

Soft X-rays are efficiently scattered at small angles by interstellar dust grains. This effect, as predicted by \citet{overbeck65} and observationally confirmed by \citet{rolf83}, is responsible for the presence of X-ray scattering halos around several bright   X-ray sources. The X-ray scattering cross section depends on
the grain properties, and the observed radial distribution of the photons
scattered in the halo also depends on the positions of the scattering grains along
the line of sight. Therefore, the study of the energy-dependent radial
profile of the X-ray halos gives useful information on the dust grain
size, composition, and spatial distribution   (see e.g. \citealt{predehl95}).

Scattered X-rays are delayed with respect to direct ones, owing to their longer path length
from the source to the observer. In the case of   variable sources, these delays
give rise to time-dependent effects on the halos properties, that can be used, if the   dust spatial distribution
is known,
to constrain the source distance \citep{ts73}. This method was applied for the first time to constrain the distance of the X-ray binary Cyg~X-3 \citep{predehl00}.

An interesting situation occurs when the dust is concentrated only in a
thin layer and the X-ray
emission has a  short duration.
In this case the halo appears as a X-ray ring with increasing radius,  because
only the photons scattered at a well determined  angle are detected at a given time.
Such ``expanding rings'' (or ``light echoes'') have been observed to date in a handful of gamma-ray bursts (GRBs)
and by measuring their expansion rate it has been possible to derive
accurate distances of Galactic dust clouds (\citealt{vianello07} and references therein).

Here we report on the discovery of dust scattering  rings around the Galactic
source \src\ \citep{tiengo09gcn8848}.
\src\ belongs to the small class of anomalous X-ray pulsars (AXPs), which, together with
the soft gamma-ray repeaters (SGRs) are thought to be magnetars, i.e. isolated
neutron stars powered by the energy stored in their extremely strong magnetic
field (see \citealt{mereghetti08} for a recent review).
These sources are characterized by the sporadic emission of very bright,
short bursts, reaching peak luminosities
above $\sim$10$^{46}$ erg s$^{-1}$ in the extreme cases known as giant flares.
In 2008 October \src\ emitted several short bursts and its X-ray
flux increased significantly \citep{israel09sub}.
No further bursts were reported until 2009 January 22, when the
source started a new period of much stronger bursting activity
\citep{mereghetti09}.
This prompted the follow-up X-ray observations that led to the discovery
of the three dust scattering rings discussed here.

The paper is organized as follows. In Section 2 we briefly review some properties of the X-ray dust scattering
process that will be used in our analysis of the
\swi\ and \xmm\  observations described in Section 3.
From the expansion rate of the three dust rings we could establish that they originate
from the scattering of the same event in three dust layers at different distances (Section 4.1). While this result is independent on the dust properties, the estimate of
the distance by modeling the rings intensity (Section 4.2)
requires the knowledge of the X-ray scattering differential cross section, hence it depends on the
assumed properties of the dust.
We found that only a subset of the different dust models that
we explored can give satisfactory fits (Section 5.1). The possible distances for
\src\ from the best-fitting models span a large range of values (Section 5.2),
but this can be restricted based on independent information
on the gas distribution in the Galaxy
(Section 5.3).

\section{X-ray dust scattering rings}

The theory of X-ray scattering by interstellar dust grains has been
reviewed in several articles (e.g. \citealt{mg86,ml91,smith98,D03}). Here we
remind only a few properties
that are relevant to our analysis of the
dust scattering rings detected around \src .

Let us consider a source at distance $d$ emitting a short burst of X-ray radiation
that is scattered by a thin layer of dust at distance $d_{\rm dust}$.
The photons scattered at larger angles reach the observer
later because of
their longer optical path. Thus the scattering halo appears as a narrow ring
centered at the source position and with angular size increasing  with time.
%
Considering that the involved angles are small (typically less than 10 arcmin),
the ring angular radius $\theta (t)$ is easily
derived by simple geometrical considerations giving:
\begin{equation}
\theta (t) = \bigg[\frac{2 c}{d} \frac{(1-x)}{x} (t-t_0)\bigg]^{1/2} \label{theta}
\end{equation}
where  $x=d_{\rm dust}/d$, $c$ is the speed of
light and  $t_0$ is the time at which unscattered photons are detected
(in the following called burst time).
The photons scattering angle  $\theta_s$ is related to the ring size by:
\begin{equation}
\theta\approx(1-x)\theta_s \label{thetacirca}\ .
\end{equation}
If the dust is in our Galaxy and the source is at cosmological distances, as in the case of GRBs, these
relations simplify to  $\theta (t) = [2 c (t-t_0)/d_{\mathrm{dust}} ]^{1/2} $ and
$\theta=\theta_s$,  thus the dust distance can be directly derived
only from the ring expansion rate.
For Galactic sources, instead, both $d$ and $d_{\rm dust}$ can be obtained
only if the parameter $x$ is independently measured.
This can be done with an analysis of the intensity evolution of the rings as
a function of the energy and of the ring radius. In fact, the scattering differential cross-section
depends on the grain size and composition, and it rapidly decreases with energy
and scattering angle (see e.g. \citealt{mg86}).
This means that,  at any given energy, the ring intensity decrease with the scattering
angle is univocally determined by the dust composition and grain size distribution.
Since the observed ring radius is
related to the scattering angle by equation [\ref{thetacirca}],
the parameter $x$, and therefore the distance of the source and of the dust layer, can be obtained by fitting the energy-resolved intensity decay of an expanding ring at different times.
This method is analogous to the analysis of the halo profiles of persistent X-ray sources (see e.g. \citealt{predehl95}), but it has the advantage that the dust distribution along the line of sight is known and the radiation scattered by each dust cloud can be analyzed separately.

For the following analysis it is useful to define an expansion coefficient $K = [19.84 (1-x)/(x d)]^{1/2}$.  Equation [\ref{theta}] can then be written as
\begin{equation}
\theta (t) = K (t-t_0)^{1/2}, \label{thetasimp}
\end{equation}
where $\theta$ is in arcmin, $t$ in days, and $d$ in kiloparsecs.
%
%

%

Assuming the Rayleigh-Gans approximation \citep{smith98},
the single-scattering halo profile (expressed in erg cm$^{-2}$ s$^{-1}$ keV$^{-1}$) predicted for a burst of fluence $F_{\rm X}$ (in erg cm$^{-2}$ keV$^{-1}$) scattered by  grains
with size distribution $n(a)$ (dust grains with radius $a$ per hydrogen atom)  is \citep{ml91}:
\begin{equation}
I(\theta,E)=3.6\times 10^{-5} \bigg(\frac{19.84+K^2d}{K~d}\bigg)^2 N_{\mathrm{H}} F_{\mathrm{X}} \int \mathrm{d} a ~ n(a)\frac{\mathrm{d}\sigma}{\mathrm{d}\Omega} \label{ithetae}
\end{equation}
where $N_{\rm H}$ is the hydrogen column density in the dust cloud, $\frac{\mathrm{d}\sigma}{\mathrm{d}\Omega}=1.1(\frac{\rho}{3})^2 a^6 \Phi^2(\theta,E,a,K,d)$ cm$^2$ is the dust scattering cross-section,
$\rho$ is the density (in g cm$^{-3}$) of each dust component and $\Phi(\theta,E,a,K,d)$ is the form factor.
In our case the Rayleigh-Gans approximation is appropriate, since in the following analysis this formula will be applied in the energy range where it is valid ($E>1.5$ keV). On the other hand, a Gaussian approximation of the form factor is often adopted, but we used the exact formula for spherical uncoated grains (equations [2.4] and [2.5] in \citealt{ml91}), since we have verified that in our case
the approximation is not
accurate enough
at large angles and at the highest energies.

Many dust models, with different compositions and grain size distributions, have been proposed and tested against multiwavelength observational data.
Weingartner \& Draine (2001, WD01) derived a model of dust composed of carbonaceous grains (polycyclic aromatic hydrocarbons at the smallest sizes and graphite for larger grains, which are the responsible for the scattering at X-rays) and amorphous silicate grains.
\citet{zda04} proposed 15 dust models, characterized by different abundances of the interstellar medium (B stars, F and G stars, or solar), by the possible presence of composite dust (in addition to bare grains) and by the form of the bare carbon (graphite, amorphous carbon, or no carbon).

Besides the WD01 model and the 15 models of \citet{zda04}, we also tried a simple
idealized model, consisting of a single component of dust grains  with a power-law  size distribution $n(a)\propto a^{-\alpha}$, defined between $a_{\rm min}$ and $a_{\rm max}$ (see e.g. \citealt{mrn77}).

\section{Observations}

The 2009 January outburst of \src\ was extensively followed by the \swi\ satellite with
the X-Ray Telescope (XRT; \citealt{burrows05}) operated
either in Windowed Timing (WT) mode, characterized by a high time
resolution but only mono-dimensional (1D) imaging, or in Photon
Counting (PC) mode, where bi-dimensional (2D) imaging of the 10$'$
radius field of view is available.
We used 5 observations in WT mode\footnote{We did not use two
WT mode observations obtained earlier (MJD~54853.303 and
54853.397) because at those times the rings were too small to
be resolved from the bright central source.} and 7 observations in
PC mode (see Table~\ref{obs}). After 2009 January 30 the
rings became too dim to be detected by the \swi/XRT instrument. We therefore
requested a 50 ks long observation with \xmm, that was performed on 2009 February 3--4\footnote{Due to
visibility constraints \xmm\ could not observe the source
earlier.} and allowed us
to detect the faint rings thanks to the great sensitivity and
large field of view of the EPIC instrument. EPIC is composed of a
PN \citep{struder01} and two MOS X-ray cameras \citep{turner01},
that in this observation were operated in Full Frame mode, with the
Thick optical blocking filter.
\begin{deluxetable}{lcccc}
  \tablecolumns{1}
\tablewidth{0pc}
\tablecaption{\label{obs} Observations of \src.}
\tablehead{
\colhead{Obs.ID} & \colhead{Start time} & \colhead{End time} & \colhead{Instrument} & \colhead{Exp./mode\tablenotemark{a}}\\
\colhead{} & \colhead{(MJD)} & \colhead{(MJD)} & \colhead{} & \colhead{(s)}
}
\startdata
00340573000 & 54853.438 & 54853.464 & \swi/XRT & 1967/WT \\
00340573001 & 54853.505 & 54853.531 & \swi/XRT & 2204/WT \\
00340573001 & 54853.572 & 54853.587 & \swi/XRT & 1288/WT \\
00340573001 & 54853.639 & 54853.656 & \swi/XRT & 1432/WT\\
00340573001 & 54853.706 & 54853.722 & \swi/XRT & 1373/WT \\
00340923000 & 54854.642 & 54854.661 & \swi/XRT & 1658/PC\\
00030956031 & 54855.262 & 54855.385 & \swi/XRT & 2560/PC \\
00341055000 & 54856.132 & 54856.259 & \swi/XRT & 3999/PC \\
00341114000 & 54856.918 & 54857.012 & \swi/XRT & 4562/PC \\
00030956032 & 54858.063 & 54858.344 & \swi/XRT & 6182/PC \\
00030956033 & 54859.210 & 54859.417 & \swi/XRT & 4508/PC \\
00030956034 & 54859.951 & 54860.216 & \swi/XRT & 6416/PC \\
0560181101 & 54865.768 & 54866.422 & \emph{XMM}/PN & 48626/FF\\
0560181101 & 54865.753 & 54866.424 & \emph{XMM}/M1 & 57040/FF\\
0560181101 & 54865.753 & 54866.424 & \emph{XMM/}M2 & 57080/FF\\
\enddata
\tablenotetext{a}{The time resolutions of the operating modes are:
XRT Windowed Timing (WT): 1.766 ms; XRT Photon Counting (PC):
2.507 s; PN Full Frame (FF): 73 ms; MOS Full Frame (FF): 2.6 s.}
\end{deluxetable}

The \swi/XRT data were processed with standard procedures using
the FTOOLS task XRTPIPELINE (version 0.11.6). We
selected events with grades \mbox{0--2} for the WT data and grades
\mbox{0--12} for the PC data. For the spectral analysis, we used
the latest available spectral redistribution matrices (version 11).
The \xmm/EPIC data were processed using the \xmm\ Science Analysis Software
(SAS version 9.0.0) and the most recent calibration files.
The standard pattern selection criteria for the EPIC X-ray
events (patterns 0--4 for PN and 0--12 for MOS) were adopted.
The \xmm\ observation was not contaminated by background flares.

\begin{figure*}
\resizebox{\hsize}{!}{\includegraphics[angle=0]{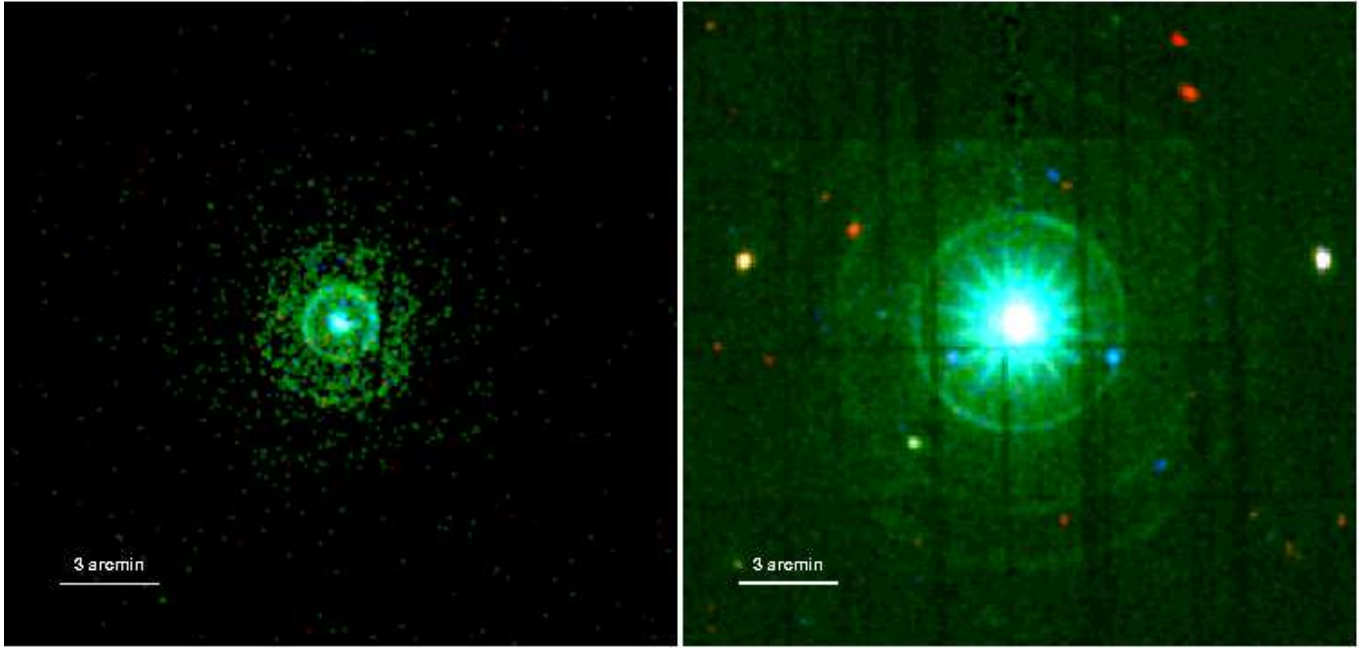}}
\caption{\label{image} The first \swi/XRT image of the rings
obtained in Photon Counting mode (left panel) and the \xmm/EPIC image
obtained 12 days later (right panel). The images are slightly smoothed and the red, green and blue colors correspond to the 0.5--1.4 keV, 1.6--4 keV and 4--7 keV energy bands, respectively.}
\end{figure*}





\section{Data analysis and results}

Figure~\ref{image} shows  the images of the first XRT observation in
PC mode and of the EPIC observation.
Three concentric rings, centered at the position of \src, are clearly
visible in the images.
They could be due to scattering
of three bursts by a thin dust layer
along the line of sight or by
three dust layers scattering the same event.

\subsection{Ring expansion}

To derive the expansion rate of the dust scattering rings we used
only the full imaging data, in which the ring radii can be well constrained
(7 XRT observations in PC mode and the sum of the three EPIC images).
We first cleaned the data by excluding time
intervals during which bright bursts  from \src\ were detected.

\begin{figure}
\centering
\resizebox{\hsize}{!}{\includegraphics[angle=0]{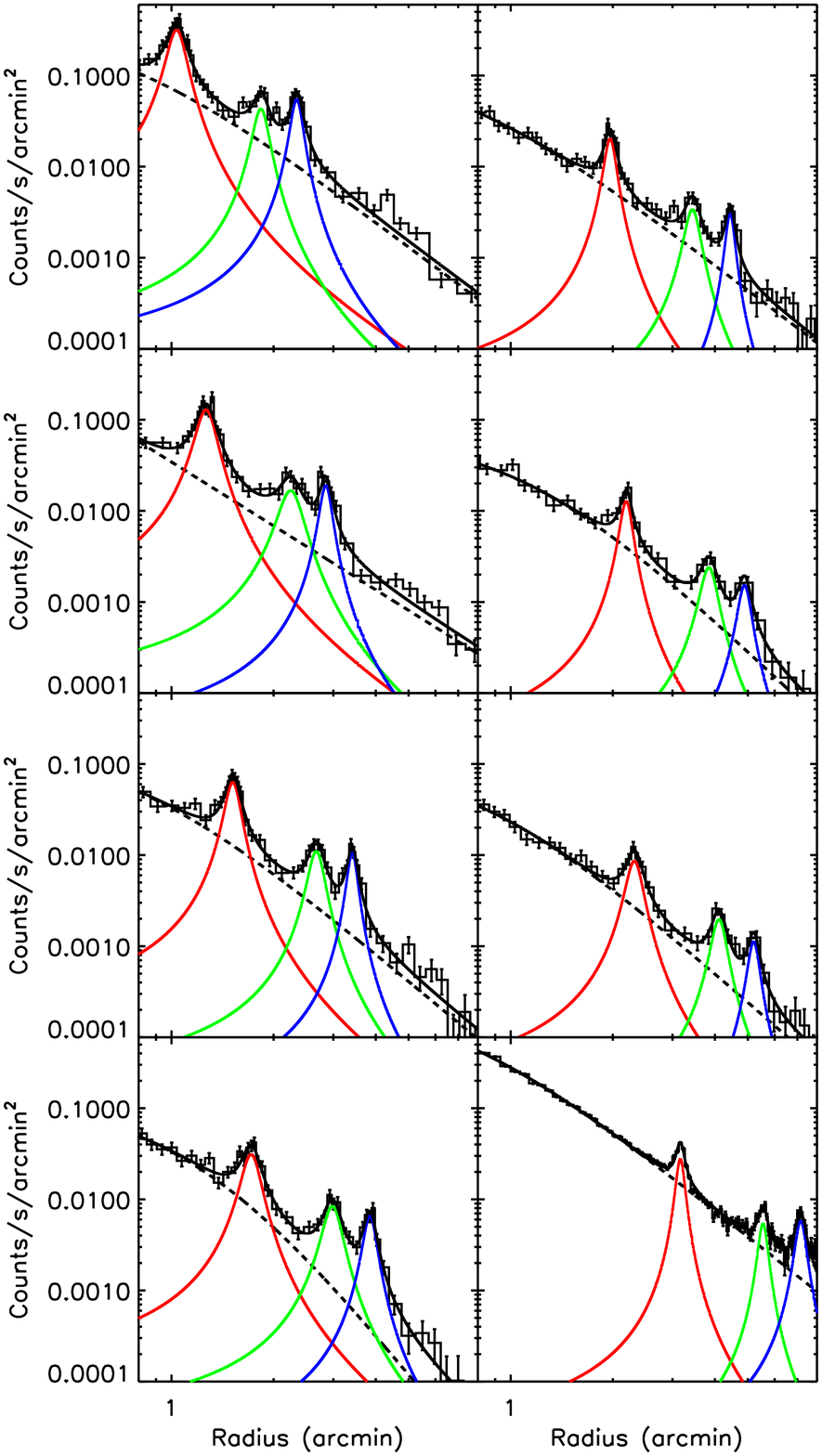}}
\caption{\label{radprof} Exposure-corrected radial profiles of the seven \swi/XRT
observations in PC mode and of the \xmm/EPIC observation. The model is
composed by a King profile (dashed line) plus three Lorentzian components (red, green, and blue). A spatially uniform background has been subtracted.}
\end{figure}

We limited the analysis of the XRT data to the 1--6 keV energy range, where the rings are significantly detected.
In order to avoid contamination by spatially variable emission lines in the instrumental background,
the analysis of the PN and MOS data was restricted to the  1.6--6 keV and 1.9--6 keV range, respectively.
In EPIC, we removed the point sources detected in the field and the
out-of-time PN events from the bright central source.
For each data set we extracted a radial profile centered at the position of \src , as
derived with a standard centroid search algorithm in each observation.
The number of counts in each radial bin was normalized to the enclosed area and divided
by the corresponding average exposure time derived from the exposure map, in order
to correct the observed profile for the vignetting, the dead detector areas, and the
excluded regions.
The resulting radial profiles, where the peaks due to three rings are clearly visible,
are shown in Figure~\ref{radprof}.

The point-spread functions (PSFs) of \swi/XRT \citep{moretti05} and
\xmm/EPIC \citep{read04} are well modeled by a King profile.
We verified through simulations that the radial profile of a thin
ring broadened by a PSF with a King profile is well fit by a
Lorentzian function. Since the rings slightly expand during each
observation, their profile is expected to be asymmetric, but
we verified that a fit with a Lorentzian  is still a good
approximation due to the relatively short duration of our observations.
Therefore we fit the radial profiles with a model consisting of
the sum of a King profile (to account for the central point
source\footnote{The derived parameters indicated also the presence
of an additional extended component;
its possible origin is discussed in Section 5.3}),
three Lorentzian components (corresponding to the three
rings) and a constant (representing the background).

\begin{figure}
\resizebox{\hsize}{!}{\includegraphics[angle=0]{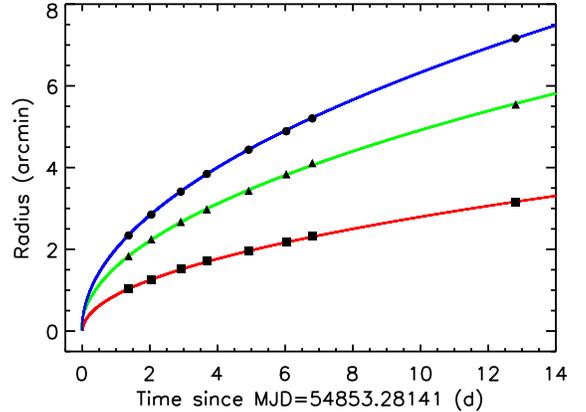}}
\caption{\label{exp} Expansion of the three dust scattering rings
as observed by \swi/XRT in PC mode and \xmm/EPIC. The lines show the
best fit  model $\theta (t) = K (t-t_0)^{1/2}  $ for the three rings. The error bars on the ring radii are not shown since they are much smaller than the symbols.}
\end{figure}

The angular size of the three rings, as obtained from the best-fit values of the
Lorentzian centroids, is plotted as a function of time in Figure~\ref{exp}.
We used equation [\ref{thetasimp}] to fit the size evolution of the three rings, with $K$ and $t_0$ as
free parameters, and obtained good $\chi^2$ values (6.2/6 degrees of freedom [dof], 5.1/6 dof and 0.8/6 dof).
The best-fit t$_0$ values for the three rings resulted consistent with a single
burst time, while three significantly different values for $K$ were found.
This means that the rings were produced by the scattering of the same burst
(or of a series of bursts occurring within a short time interval) by three dust
layers at different distances.

The derived t$_0$ values are compatible with the short time
interval from 6:43 to 6:51 UT of 2009 January 22, in which the highest bursting activity
was seen during a long, uninterrupted observation obtained with
the Anti-Coincidence Shield (ACS) of the Spectrometer on \integral\ (SPI)
instrument \citep{mereghetti09}.
Among the several tens of short bursts seen in this interval, two events were outstanding
for their exceptional brightness: a burst at 6:45:14 UT lasting $\sim$1.5 s
and a longer event that occurred at  6:48:04 UT.
The latter burst consisted of an initial
bright spike, lasting 0.3 s, followed by a tail pulsed at the 2.1 s rotation period of \src.\footnote{The other pulsed  burst detected at 8:17:29 UT \citep{mereghetti09} is incompatible with the possible values for $t_0$
and therefore cannot be the origin of the X-ray rings.}
Both events saturated the ACS and the first one saturated also the detector on
the \emph{RHESSI} satellite \citep{bellm09}. Therefore it is not possible to
establish which of the two bursts had a higher fluence in the soft X-ray range
and caused the scattering rings.  Due to the small time delay between  these two events,
all our analysis and results of the next section do not depend significantly
on which of these burst was the origin of the rings
(in fact it is also possible that both
events contributed to the observed scattered X-rays).
In the following we will assume that the
observed rings were produced by the first burst and define  $T_0 = 54853.28141$ MJD = 2009 January 22 at 6:45:14 UT.
The fits result gave $t_{0,1}=T_0-1000\pm1100$ s, $t_{0,2}=T_0-4000\pm1900$ s,
and $t_{0,3}=T_0+300^{+1100}_{-1600}$ s (all the errors are at 1 $\sigma$).
Fitting jointly the three rings  imposing a common $t_0$,
we obtain a 90\% confidence interval $T_0-2100$ s $<t_0<T_0+700$ s.
Finally, by fixing $t_0=T_0$, the fit yields
the  expansion coefficients
$K_1=0.8845\pm0.0008$,
$K_2=1.553\pm0.003$,
and $K_3=2.000\pm0.002$ ($\chi^2$/dof=17.2/21).


%


\subsection{Spectra and radial profiles}

Here we describe our analysis of the energy-dependent halo profiles, i.e.
the spectra $I(E,\theta)$ of the three rings at the different angles  $\theta$ from the
source direction.
In fact, having already determined the expansion
rate, for any given dust scattering model,  $I(E,\theta)$ depends only on the source distance, column density $N_{\rm H}$,  and fluence $F_{\mathrm{X}}(E)$ of the primary radiation (see equation \ref{ithetae}).
By simultaneously fitting $I(E,\theta)$ for the three rings we will derive some
information on these quantities.

While the relative optical depth of the three dust layers is set by the relative
intensity of the three rings, their absolute values are unknown because the
burst fluence $F_{\mathrm{X}}(E)$ in the soft X-ray band was not directly measured.
In the following analysis we fixed the column density of the dust cloud producing
the innermost and brightest ring at  $N_{\rm H1}=10^{22}$ cm$^{-2}$.
This value was chosen considering that the total $N_{\rm H}$ along the line of sight derived from the X-ray spectrum
of \src\ is $\sim$$3\times10^{22}$ cm$^{-2}$ \citep{halpern08}.
The burst fluence $F_{\mathrm{X}}(E)$ that we will derive from the fits of $I(E,\theta)$
depends on the assumed value $N_{\rm H1}$. Also the column densities obtained for the two other
clouds, $N_{\rm H2}$ and $N_{\rm H3}$, will be relative to  $N_{\rm H1}$.
However, it is important to note that the derived distances  do not depend on $N_{\rm H1}$ since, for any fixed dust model, they are determined only by the
relative intensity of the  rings at different times and energies (see equation [3]).

The extraction of the ring spectra by selecting annular regions
(in the data sets with 2D images) is hampered by the spatial proximity
of the three rings and by the contamination from the bright
central point source and from the diffuse halo. Therefore, we
followed a different approach to obtain the background-subtracted
spectra of the rings: we derived the number of
counts per energy bin by fitting the energy resolved radial
profiles of every observation.
In order to properly take into account for all the instrumental and systematic
effects, we did not use an analytic expression (as done in the
previous section), but we resorted to detailed simulations to derive the expected radial profile of each of the following five components: the three rings, the central source and the background.

For each observation listed in Table~\ref{obs} and for each component separately,
we simulated a 2-D image taking into account the effect of the instrumental PSF.
A sufficiently large number of photons was generated in order to minimize statistical uncertainties. The rings expansion was simulated based on the real observation plan and on the expansion law
derived above.\footnote{The two outer rings in the EPIC data are significantly broader than predicted by the PSF and the ring expansion. We could adequately model their profiles  adopting a 200 pc thickness for the corresponding dust layers. Although this effect is detectable only in the EPIC images, we introduced it also in the simulations of the two outer rings for all the \swi\ observations.}
The resulting images were convolved with the exposure maps of the
corresponding real observations, in order to correct for the vignetting, detector
defects and gaps and, in the case of the \xmm\ data, the regions excluded to
eliminate the field point sources (and out-of-time events for the PN).

At this point, we produced the radial profiles of the simulated XRT/PC and EPIC images,
while for the simulations of the 5 XRT observations in WT mode we projected the images onto the X axis of the XRT CCD. These
histograms, with a variable normalization factor,
were used to fit
the radial profiles and the 1-D WT images observed at the different energies.
The ratio between the number of photons in the simulation and the normalization factor
(with the statistical error obtained by the fitting procedure) provides the
number of photons of the component in the selected energy range.
Using this method to extract all the spectra, no PSF correction must
be applied to the ancillary response files (ARF), because the PSF
profile is integrated over the whole detector plane. Similarly, no
corrections for vignetting and chip defects must be applied, because they were already taken into account during the simulation process.
The WT spectra of the two outermost rings were excluded from the following
analysis because their counts could not be unambiguously separated in the WT profiles.

To properly model the observed X-ray profiles, the addition of a diffuse halo component
was also required.
Since it has a non-uniform spatial distribution that we are not able to model
{\it a priori}, we described  it with an analytical function convolved with the exposure map: for the 2-D data we used  a King profile centered on \src , while for the WT data we used a King function convolved with a constant gradient along the X axis
because we detected an asymmetry of the diffuse halo.

\begin{figure}
\centering
\resizebox{\hsize}{!}{\includegraphics[angle=0]{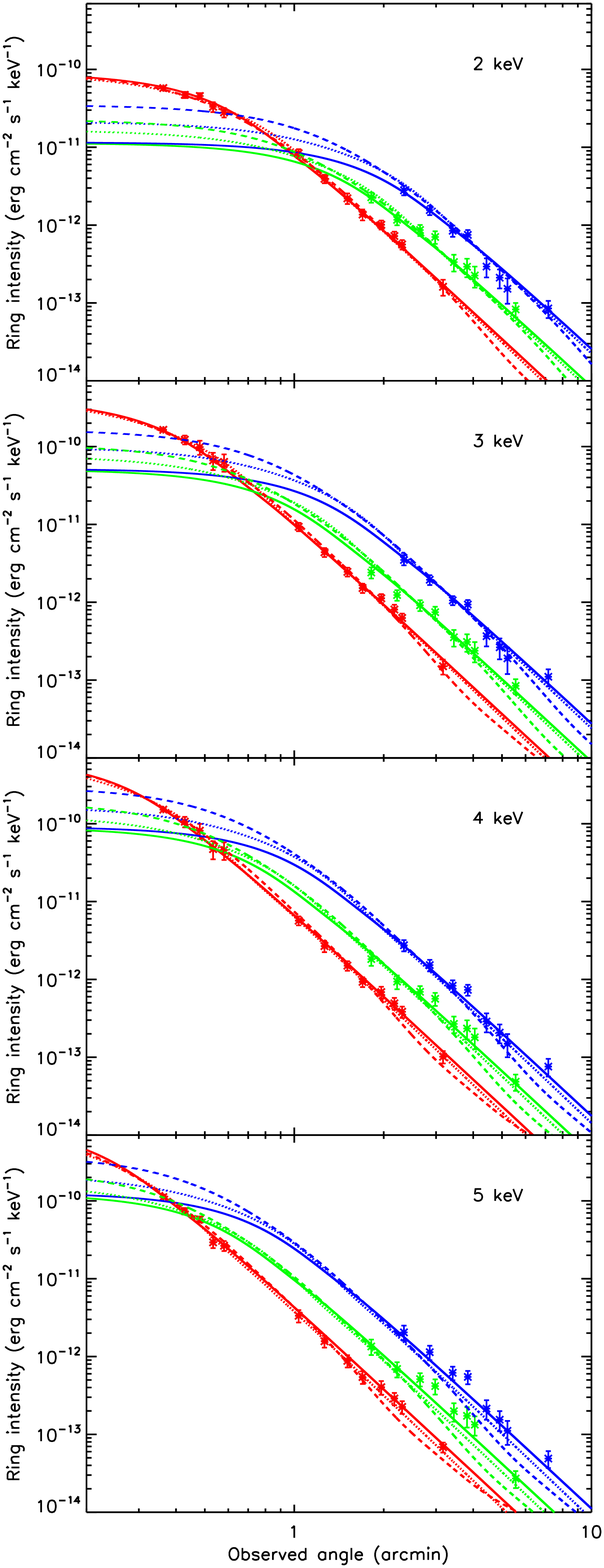}}
\caption{\label{figdust} Halo profiles at different energies for the three rings. The models are based on the following dust compositions and grain size distributions:
WD01 from \citet{WD01} (dotted lines), BARE-GR-B (solid lines) and COMP-NC-FG  (dashed lines) from \citet{zda04}. See Table~\ref{dust} for the corresponding best-fit parameters.}
\end{figure}

With the procedure described above we obtained the rings spectra for each
observation. We then fitted these spectra with a phenomenological model (an absorbed  power-law)
in order to obtain the ring's specific intensities as
a function of $\theta$ at seven energy values between 2 and 7 keV.
As an example  we show in Figure~\ref{figdust} the results for 2, 3, 4 and 5 keV,
where the data points of the three rings are indicated by different colors.
The 21 radial profiles (seven energies times three rings) were simultaneously
fitted with the model of equation [\ref{ithetae}] for the different dust compositions and grain size distributions described in Section 2, yielding the results reported in
Table~\ref{dust}. A joint fit is justified by the assumption that the three dust clouds are spatially uniform in the plane of the sky and are made of dust with the same composition and grain size distribution.
\begin{deluxetable}{lcccc}
  \tablecolumns{1}
\tablewidth{0pc}
\tablecaption{\label{dust} Best-fit parameters of the halo profiles  with different dust models.}
\tablehead{
\colhead{Dust model} & \colhead{Distance} & \colhead{$\frac{N_{\rm H2}}{N_{\rm H1}}$}& \colhead{$\frac{N_{\rm H3}}{N_{\rm H1}}$} & \colhead{$\chi^2_{\rm red}$/dof}\\
\colhead{} & \colhead{(kpc)} & \colhead{} & \colhead{} & \colhead{}
}
\startdata
bare-gr-b&  3.91$\pm$0.07& 0.24$\pm$0.01& 0.27$\pm$0.01& 0.77$/$193\\
bare-gr-s&  4.76$\pm$0.08& 0.25$\pm$0.01& 0.30$\pm$0.01& 0.79$/$193\\
bare-gr-fg&  4.86$\pm$0.09& 0.25$\pm$0.01& 0.30$\pm$0.01& 0.81$/$193\\
comp-gr-b&  5.22$\pm$0.10& 0.26$\pm$0.01& 0.31$\pm$0.01& 0.85$/$193\\
power-law\tablenotemark{a} &	5.40$\pm$0.13	& 0.26$\pm$0.01	& 0.32$\pm$0.01			&	 0.86$/$192\\
comp-gr-fg&  6.95$\pm$0.13& 0.28$\pm$0.01& 0.35$\pm$0.01& 1.05$/$193\\
power-law\tablenotemark{b} &	4.86$\pm$0.08	& 0.25$\pm$0.01	& 0.29$\pm$0.01			&	 1.09$/$193\\
comp-gr-s&  7.71$\pm$0.14& 0.29$\pm$0.01& 0.37$\pm$0.01& 1.10$/$193\\
power-law\tablenotemark{c} &	6.05$\pm$0.10	& 0.27$\pm$0.01	& 0.33$\pm$0.01			&	 1.12$/$193\\
power-law\tablenotemark{d} &	7.33$\pm$0.13	& 0.28$\pm$0.01	& 0.36$\pm$0.01			&	 1.14$/$193\\
WD01 &  6.91$\pm$0.12& 0.27$\pm$0.01& 0.34$\pm$0.01& 1.27$/$193\\
comp-nc-b& 11.83$\pm$0.24& 0.33$\pm$0.01& 0.46$\pm$0.01& 1.33$/$193\\
bare-ac-s&  5.74$\pm$0.10& 0.26$\pm$0.01& 0.31$\pm$0.01& 1.36$/$193\\
bare-ac-b&  4.83$\pm$0.08& 0.24$\pm$0.01& 0.28$\pm$0.01& 1.37$/$193\\
bare-ac-fg&  5.85$\pm$0.10& 0.26$\pm$0.01& 0.31$\pm$0.01& 1.44$/$193\\
comp-ac-b&  8.37$\pm$0.16& 0.27$\pm$0.01& 0.36$\pm$0.01& 1.52$/$193\\
comp-ac-s&  9.24$\pm$0.17& 0.29$\pm$0.01& 0.39$\pm$0.01& 1.55$/$193\\
comp-ac-fg&  8.14$\pm$0.15& 0.28$\pm$0.01& 0.36$\pm$0.01& 1.73$/$193\\
comp-nc-s& 10.17$\pm$0.19& 0.29$\pm$0.01& 0.40$\pm$0.02& 1.81$/$193\\
comp-nc-fg& 10.36$\pm$0.24& 0.30$\pm$0.01& 0.41$\pm$0.01& 1.87$/$193\\
\enddata
\tablenotetext{a}{Power-law grain size distribution $a^{-\alpha}$ with $\alpha=3.66\pm0.02$, $a_{\mathrm{min}}=0.0003$ $\mu$m and $a_{\mathrm{max}}=0.3$ $\mu$m.}
\tablenotetext{b}{Power-law grain size distribution $a^{-\alpha}$ with $\alpha=3.5$, $a_{\mathrm{min}}=0.0003$ $\mu$m and $a_{\mathrm{max}}=0.25$ $\mu$m.}
\tablenotetext{c}{Power-law grain size distribution $a^{-\alpha}$ with $\alpha=3.5$, $a_{\mathrm{min}}=0.0003$ $\mu$m and $a_{\mathrm{max}}=0.3$ $\mu$m.}
\tablenotetext{d}{Power-law grain size distribution $a^{-\alpha}$ with $\alpha=3.5$, $a_{\mathrm{min}}=0.0003$ $\mu$m and $a_{\mathrm{max}}=0.35$ $\mu$m.}
\end{deluxetable}

It is evident that, although formally acceptable fits were obtained with different
dust models and the corresponding errors on the fitted parameters are
rather small, there is a wide spread in the derived distances for \src.\footnote{In order to discriminate among these dust models, an independent measure of the distance of the X-ray source or of at least one of the three dust clouds would be required. A robust measure of any of these distances is still lacking, but the currently available constraints are discussed in the next section.}
All the dust models give reasonably similar
values for the relative $N_{\rm H}$ values of the three dust clouds.

\section{Discussion}

By measuring the expansion rate of the dust scattering rings seen in X-ray
observations of the AXP  \src\ we  established that they were produced by three dust clouds at different distances, and that the most likely origin of the scattered X-rays was
the extremely bright burst that occurred on 2009 January 22 at 6:45:14 UT   \citep{mereghetti09}.
With a detailed spatial and spectral analysis of the three X-ray rings we could
test different dust models and derive some information on the distance of \src ,
on the distance and optical depth of the three dust clouds, and on the $\sim$2--7 keV
properties of the burst (that was observed directly only at higher energy).
Despite the large systematic uncertainty on the \src\ distance, resulting from  our
ignorance of the actual dust properties, our analysis yielded some
results that have implication both for the study of the Galactic dust properties and
of \src .

\subsection{Dust models}

The   dust models that we adopted are characterized by  different   cross
sections for X-ray scattering. As a consequence the quality of the fits to the
energy-dependent angular profiles of the rings is not the same in all
models, as demonstrated by the $\chi^2$ values reported in Table~\ref{dust}.
It is remarkable that the models giving the best fits, namely the ones containing bare graphite grains and no composite dust, are the same already found to better describe the dust-scattering halos of other Galactic X-ray sources \citep{smith06,smith08,valencic09}. On the other hand, the models with no carbon or amorphous carbonaceous grains provide significantly poorer fit to our X-ray data.
Within the same class of models, we generally find a better fit for the dust models constructed according to the abundances of B stars.
However, since the profiles of the X-ray scattering halos are more sensitive to the dust grain size distribution rather than to its chemical composition, the elemental abundances of the interstellar medium cannot be directly constrained by our analysis.

As can be seen from the examples shown in Figure~\ref{figdust}, the models yielding
unacceptable fits systematically underestimate the ring flux
at high energies and large off-axis angles.
This means that such models have a deficit of relatively small grains ($a\sim0.01$--0.03 $\mu$m), which  are the most efficient in scattering the harder X-rays and are responsible for the larger   scattering angles.

\subsection{Distance of \src\ and burst energy}

The distances of \src\ derived assuming the different dust models have very small statistical errors, but span a large range of values.
In fact the source distance is particularly  sensitive to the dimensions of the largest
grains, which are  mainly responsible for the  scattering at the smallest angles of soft X-rays. The largest grains determine the ring fluxes at low energies during the first observations, where we detected most of the scattered photons.
The strong dependence of the source distance on the size of the largest dust grains is
well demonstrated by our fits with a power-law grain size distribution
(see Table~\ref{dust} for a few examples). Keeping the same power-law index and minimum
grain dimension, we found that the source distance is related to the maximum grain
radius by  $d\propto a_{\mathrm{max}}$$^{4/3}$.

All the dust models giving the best fits to the halo profiles yield
source distances in the relatively narrow range $d\sim4$--5 kpc, and, as noted above,  they are the most successful 
also
for other Galactic X-ray sources.
This distance estimate for \src\ agrees with that suggested on   the basis of its possible association with the supernova remnant  G\,327.24--0.13 and a star forming region \citep{gelfand07}.
A distance of 9 kpc was instead derived   from the dispersion
measure of the \src\ radio pulses  \citep{camilo07}, assuming the Galactic free electron density model of \citet{cordes02}. This model is known to have an average uncertainty of $\sim$30\%, which, however, can be much larger for individual objects.


The results of our fit  give also some constraints on the X-ray properties
of the burst that produced the scattering rings. In principle,
once the  column density in one cloud is fixed, the burst spectrum could be obtained
directly from the normalizations of the models fitting the energy-resolved halo profiles. However, due to the limited energy band and the relatively large normalization errors, the broad-band spectral shape of the burst is poorly constrained.
Therefore, we adopted for the burst an absorbed bremsstrahlung model with temperature
fixed at 30 keV or 100 keV,
as suggested by the observations of other magnetar bursts.
The corresponding best-fit parameters found for the different dust models are reported in Table~\ref{prompt}. The derived fluences vary by a factor $\sim$3, because the dust
scattering efficiency changes in the different models.
%
The table reports also the burst energy in the 1--100 keV energy band,
computed with the appropriate source distance resulting from  each dust model.
The derived values are in the range $\sim$10$^{44-45}$ erg, i.e. not far from
the energy emitted in the giant flares of SGR\,0526--66 \citep{mazets79} and SGR\,1900+14 \citep{hurley99}.
However, we remind that our energy estimate is inversely proportional to the value for the
dust optical depth that we have assumed by fixing $N_{\rm H1}$=10$^{22}$ cm$^{-2}$.

\begin{deluxetable}{lccccc}
  \tablecolumns{1}
\tablewidth{0pc}
\tablecaption{\label{prompt} Best-fit parameters with an absorbed bremsstrahlung model of the burst unscattered spectrum.}
\tablehead{
\colhead{Dust model} & \colhead{$N_{\rm H}$} & \colhead{$kT$\tablenotemark{a}} & \colhead{$F_{burst}$\tablenotemark{b}} & \colhead{$E_{burst}$\tablenotemark{c}} & \colhead{$\chi^2_{\rm red}$\tablenotemark{d}}\\
\colhead{} & \colhead{(cm$^{-2}$)} & \colhead{(keV)}  & \colhead{(erg cm$^{-2}$)} & \colhead{(erg)} & \colhead{}
}
\startdata
bare-gr-b & 4.5$\times$10$^{22}$  & 30 & 0.0050  & 1.0$\times$10$^{44}$  & 1.02 \\
 & 4.2$\times$10$^{22}$ & 100 & 0.0050 & 1.9$\times$10$^{44}$ & 0.17 \\
bare-gr-s & 4.5$\times$10$^{22}$ & 30 & 0.0039 & 1.2$\times$10$^{44}$ & 0.99 \\
 & 4.2$\times$10$^{22}$ & 100 & 0.0039 & 2.2$\times$10$^{44}$ & 0.17 \\
bare-gr-fg & 4.5$\times$10$^{22}$ & 30 & 0.0038 & 1.2$\times$10$^{44}$ & 1.27 \\
 & 4.2$\times$10$^{22}$ & 100 & 0.0038 & 2.2$\times$10$^{44}$ & 0.30 \\
comp-gr-b & 4.4$\times$10$^{22}$ & 30 & 0.0045 & 1.6$\times$10$^{44}$ & 0.25 \\
 & 4.2$\times$10$^{22}$ & 100 & 0.0045 & 3.0$\times$10$^{44}$ & 0.23 \\
comp-gr-fg & 4.4$\times$10$^{22}$ & 30 & 0.0034 & 2.2$\times$10$^{44}$ & 0.23 \\
 & 4.2$\times$10$^{22}$ & 100 & 0.0034 & 4.0$\times$10$^{44}$ & 0.70 \\
comp-gr-s & 4.4$\times$10$^{22}$ & 30 & 0.0033 & 2.6$\times$10$^{44}$ & 0.31 \\
 & 4.2$\times$10$^{22}$ & 100 & 0.0033 & 4.8$\times$10$^{44}$ & 1.05 \\
WD01 & 4.3$\times$10$^{22}$ & 30 & 0.0020 & 1.3$\times$10$^{44}$ & 0.70 \\
 & 4.1$\times$10$^{22}$ & 100 & 0.0020 & 2.3$\times$10$^{44}$ & 1.82 \\
comp-nc-b & 4.5$\times$10$^{22}$ & 30 & 0.0059 & 1.1$\times$10$^{45}$ & 0.96 \\
 & 4.2$\times$10$^{22}$ & 100 & 0.0059 & 2.0$\times$10$^{45}$ & 0.33 \\
bare-ac-s & 4.3$\times$10$^{22}$ & 30 & 0.0036 & 1.6$\times$10$^{44}$ & 0.39 \\
 & 4.1$\times$10$^{22}$ & 100 & 0.0036 & 2.9$\times$10$^{44}$ & 0.64 \\
bare-ac-b & 4.3$\times$10$^{22}$ & 30 & 0.0045 & 1.4$\times$10$^{44}$ & 0.35 \\
 & 4.1$\times$10$^{22}$ & 100 & 0.0045 & 2.6$\times$10$^{44}$ & 0.76 \\
bare-ac-fg & 4.3$\times$10$^{22}$ & 30 & 0.0035 & 1.6$\times$10$^{44}$ & 0.45 \\
 & 4.1$\times$10$^{22}$ & 100 & 0.0035 & 2.9$\times$10$^{44}$ & 0.83 \\
comp-ac-b & 4.8$\times$10$^{22}$ & 30 & 0.0077 & 7.6$\times$10$^{44}$ & 1.93 \\
 & 4.5$\times$10$^{22}$ & 100 & 0.0077 & 1.4$\times$10$^{45}$ & 0.59 \\
comp-ac-s & 4.5$\times$10$^{22}$ & 30 & 0.0041 & 4.8$\times$10$^{44}$ & 0.33 \\
 & 4.3$\times$10$^{22}$ & 100 & 0.0041 & 8.7$\times$10$^{44}$ & 1.15 \\
comp-ac-fg & 4.4$\times$10$^{22}$ & 30 & 0.0042 & 3.8$\times$10$^{44}$ & 0.43 \\
 & 4.1$\times$10$^{22}$ & 100 & 0.0042 & 6.8$\times$10$^{44}$ & 1.39 \\
comp-nc-s & 4.6$\times$10$^{22}$ & 30 & 0.0050 & 7.2$\times$10$^{44}$ & 0.31 \\
 & 4.4$\times$10$^{22}$ & 100 & 0.0050 & 1.3$\times$10$^{45}$ & 0.28 \\
comp-nc-fg & 4.6$\times$10$^{22}$ & 30 & 0.0050 & 7.4$\times$10$^{44}$ & 0.25 \\
 & 4.3$\times$10$^{22}$ & 100 & 0.0050 & 1.3$\times$10$^{45}$ & 0.54 \\
\enddata
\tablenotetext{a}{Fixed.}
\tablenotetext{b}{Observed fluence in the 2--5 keV energy band, assuming $N_{\rm H1}=10^{22}$ cm$^{-2}$.}
\tablenotetext{c}{Unabsorbed energy emitted in the 1--100 keV energy band, assuming $N_{\rm H1}=10^{22}$ cm$^{-2}$ and the source distances reported in Table~\ref{dust}.}
\tablenotetext{d}{For 5 dof.}
\end{deluxetable}

\subsection{Distances and properties of the dust clouds}

The distances of the three dust clouds are related to the expansion coefficients $K$ and
the source distance $d$ by $d_{\mathrm{dust}}=19.84$ $d /(19.84 + K^2 d)$ kpc.
Therefore, for each model of Table~\ref{dust} we can  derive the
distances of the three clouds and compare them with independent estimates of the
dust distribution in the direction of \src .

\begin{figure}
 \begin{minipage}[t]{\hsize}
\resizebox{\hsize}{!}{\includegraphics[angle=0]{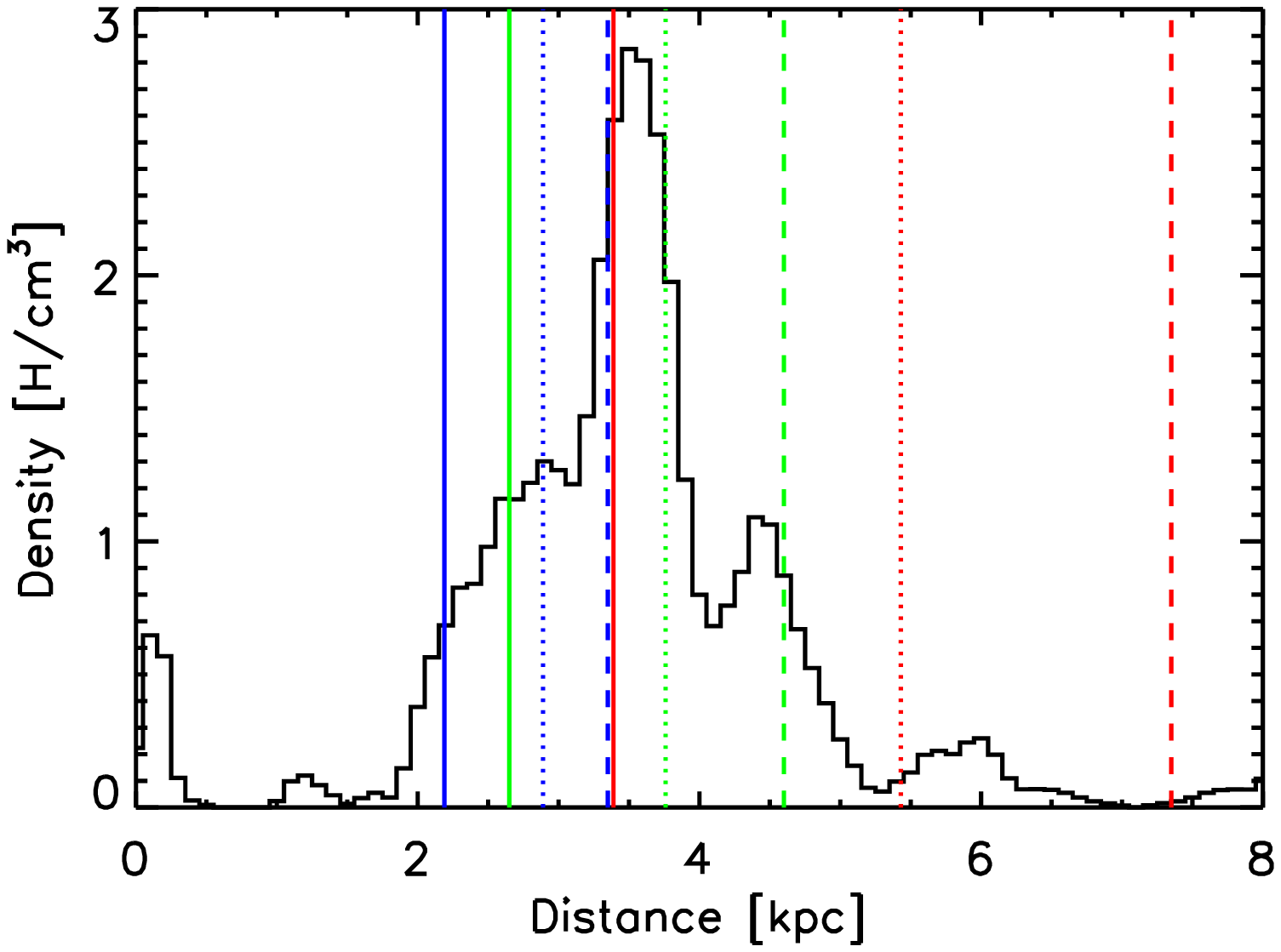}}
\caption{\label{co} Density of molecular hydrogen in the direction of \src, as derived from CO observations. The vertical lines show the inferred distances of the three dust clouds (red, green and blue for the clouds generating the smallest, intermediate and largest ring, respectively) for the following dust models: WD01 from \citet{WD01} (dotted lines), BARE-GR-B (solid lines) and COMP-NC-FG  (dashed lines) from \citet{zda04}. See Table~\ref{dust} for the corresponding best-fit parameters.}
\end{minipage}
\end{figure}

In order to obtain the distribution of molecular hydrogen along the
line of sight we use the CO observations obtained by the
Millimeter-Wave Telescope at the Harvard-Smithsonian Center for Astrophysics (CfA) \citep{dame01}. The CO is
assumed to be a tracer of molecular hydrogen, through a known ratio
($X$) between hydrogen density and CO radio emissivity; for this work we
assumed $X=2.3\times 10^{20}$ cm$^{-2}$ K$^{-1}$ km$^{-1}$ s.
The velocity-resolved radio data have been deprojected using the
Galactic rotation curve parameterized by \citet{Clemens85}.
 As shown in Figure~\ref{co}, the strongest peak of CO
emission is located at a distance of $\sim$3.5 kpc.
This value is consistent with the distance of the farthest  cloud
(producing the innermost and brightest ring) if \src\ is at $\sim$4 kpc, as found
with the models that gave the lowest $\chi^2$ in the halo profile fits. In particular, the best fitting model BARE-GR-B gives  $d=3.9$ kpc   and $d_{\rm dust} = 3.4$, 2.6 and 2.2 kpc
for the three clouds (see solid lines in Figure~\ref{co}). The presence of a large amount of dust at a distance of $\sim$3.5 kpc in this  direction is also confirmed by the analysis of the infrared extinction of the field stars   \citep{marshall06}.

The relative amount of dust in the three clouds can be inferred by the ratio of their hydrogen column densities, since, for each dust model, we used the same fraction of dust grains per hydrogen atom in the three clouds. All the dust models give similar column density ratios (see Table~\ref{dust}), with the farthest cloud containing $\sim$2--4
times more dust than each of the other two clouds.
This scenario is also compatible with the CO data and a source distance of $\sim$4 kpc, since the two closest clouds could be the responsible for the lower and broader peak of CO emission detected at a distance of 2--3 kpc (see Figure~\ref{co}).
For larger \src\ distances  the largest CO emission peak could  be
associated with one of the closer dust  clouds. However,  in this case the cloud
responsible for the brightest ring would be located at a larger distance, where, based on CO emission data, a smaller amount of dust is expected.
In conclusion, it seems that also from an analysis of the plausible locations
of the three dust clouds, a distance of $\sim$4 kpc for \src\ is favored.

It is reasonable to expect that not all the dust along the line of sight
be concentrated in the three clouds responsible for the rings.
This is supported by the analysis of the radial profiles discussed in Section 4.2.
In fact, in all observations, we obtained best-fit parameters of the
King profile different from those expected for a point source,
clearly indicating the presence of a further component of diffuse
emission around \src. We found that the intensity of this
component significantly decreases with time,  consistently with its interpretation
as scattering of the \src\ burst by dust distributed along the whole line of sight.
Another possibility is  multiple scattering by the three thin dust
layers; moreover, some contribution to this halo can come from the
dust scattering of the X-ray emission of the other bursts
occurring during the source activation and of the persistent X-ray
emission of \src .

We finally note that the diffuse X-ray emission recently reported by \citet{vink09} and interpreted as  emission of a pulsar wind nebula powered by \src\ and of the supernova remnant G\,327.24--0.13 is significantly dimmer than the extended emission (either the expanding rings or diffuse halo) described here.

\section{Conclusions}

Three bright X-ray rings were discovered around the anomalous X-ray pulsar \src\ after its
strong bursting activity of 2009 January 22.
Their radii were seen to increase and their flux to decrease with time, as expected for a short burst of X-rays scattered by three thin dust layers.
A similar phenomenon had been observed in a few GRBs, but on a much smaller scale: in the best case, the rings around GRB\,031203 \citep{vaughan04} were observed by \xmm\ from 5 to 20 hours after burst, but only $\sim$3,000 ring photons could be collected.
The $\sim$65,000 ring photons we detected for \src\ allowed us to carry out
a detailed analysis in which several subtle effects, that could have been neglected
with a smaller statistics, had to be carefully taken into account.

Based only on simple, model-independent geometrical considerations,
we found that the delayed X-rays seen in the rings were emitted by \src\ within the same
time interval of the unscattered radiation detected between 6:10 and 6:56 UT
of 2009 January 22 in hard X-rays.
This short time period includes the phase of highest bursting activity ever observed
from this source.

%

Interesting information on \src\ and on the dust clouds producing the X-ray rings
could be derived from the analysis of the ring's spectra and intensity evolution.
However, this analysis requires the knowledge of the interstellar dust properties
affecting the X-ray scattering cross section. We therefore explored the
consequences of using different dust models.

The energy released by the event that formed the X-ray scattering rings could be derived from the fit of the halo profiles, that provides at the same time its spectrum and fluence in the 2--5 keV band, and the source distance.
Assuming a column density of 10$^{22}$ cm$^{-2}$ for the most distant cloud
and a bremsstrahlung spectrum with $kT=30$--100 keV, a burst energy between 10$^{44}$ erg and 2$\times$10$^{45}$ erg was derived (in the 1--100 keV range).
These values are significantly larger than that  in the pulsating tail detected on 2009 January 22 at 6:48 UT ($E\sim2.4\times10^{43}d^2_{10~{\rm kpc}}$ erg; \citealt{mereghetti09}), that therefore cannot account for the rings.
More likely candidates are two bright bursts, whose fluence could not be determined
because they were so bright to saturate the detectors, and occurred at times
compatible with the expansion rate measured for the rings.
However, we cannot exclude that the event producing the scattering rings had a very steep spectrum and so dominated the emission at soft X-rays without showing up as an exceptional event at higher energies.


According to our estimate, the energy released by the event that generated the X-ray scattering rings is comparable to that emitted by the giant flares of SGR\,0526--66 \citep{mazets79} and SGR\,1900+14 \citep{hurley99}.
These events were followed by  pulsating tails lasting several minutes,
contrary to the case of \src: the only pulsating tail detected by the continuous \integral\ ACS observation covering the time interval compatible with the origin of the X-ray rings lasted only 8 seconds
\citep{mereghetti09}.
On the other hand, in case the X-ray scattering rings of \src\ were produced by a single short burst, it must have been exceptionally bright, since no other magnetar bursts without any pulsating tail and an emitted energy $>$10$^{44}$ erg have ever been detected. The intermediate flare emitted by SGR\,1627--41 on 1998 June 18 \citep{mazets99} was the former record holder, with $E\sim10^{43}$ erg (assuming a distance of 11 kpc).
We caution that large systematic uncertainties affect our estimate of the emitted energy: the total scattering cross-section and source distance depend on the dust model, the hydrogen column density and dust-to-gas ratio in the dust clouds are only poorly constrained, and the broad band spectrum is inferred from a narrow band X-ray spectrum and in analogy to similar events.

By using several dust models discussed in
the literature we showed that only about half of them  provide an adequate fit to the three halo profiles at different energies.
The best-fitting model (BARE-GR-B in \citealt{zda04})  yields for \src\ a distance of 4 kpc, consistent with the value derived from  its association
with the SNR  G\,327.24--0.13 \citep{gelfand07}.
This model also gives distances for the three dust clouds that are consistent
with CO observations in this direction.

Our best fitting dust model is the same that provided  the best fits to the halo
of GX5--1 \citep{smith06},
but also other models, that were compatible with the analysis of X-ray halos of other X-ray binaries (see, e.g., \citealt{smith08}),
gave good fits to our data. They resulted in distances for \src\ in the 4.8--5.4 kpc range.
%
%
%
Other dust models giving relatively good fits imply distances of 6--8 kpc,  but
require that the highest peak seen in the CO emission corresponds to the cloud
responsible for the second X-ray ring. This is not the brightest ring,
and thus these scenarios would imply  that the strength of the CO line
is not proportional to the amount of dust that efficiently scatters X-rays.

We finally note  that, if the X-ray source and dust distance are not independently constrained, the quality of the fits of the X-ray halo profiles is
mostly sensitive to the grain size distribution in the range 0.01--0.03 $\mu$m, while the source distance is mainly determined by the largest grains.
These grains, with size $>$0.1 $\mu$m, are those less constrained by diagnostics at longer wavelengths. Therefore, we cannot exclude that the actual dust properties may be different
from all the models we tested, providing a good fit to the X-ray rings, but for a significantly different distance of \src.

\acknowledgements
This research is based on observations with the NASA/UK/ASI \swi\
mission and with \xmm, an ESA science mission with instruments and
contributions directly funded by ESA Member States and NASA. We
thank the \swi\ and \xmm\ duty scientists and mission planners for
making these target of opportunity observations possible. We thank T.M. Dame for granting us access to the CO emission data in the direction of \src. We thank E. Bellm and J. Ripa for useful discussion of RHESSI results. We acknowledge the partial
support from ASI (ASI/INAF contracts I/088/06/0 and AAE~TH-058).
PE thanks the Osio Sotto city council for support with a
G.~Petrocchi fellowship. SZ acknowledges support from STFC. NR is
supported by a Ram\'on~y~Cajal fellowship. DG acknowledges the CNES for
financial support.

{\it Facilities:} \facility{{Swift (XRT)}}, \facility{{XMM-Newton (EPIC)}}

\end{document}